\pdfoutput=1
\documentclass[review,authoryear]{elsarticle}
\usepackage{color}
\usepackage{hyperref}
\hypersetup{
	colorlinks=true,
		citecolor=red,
	linkcolor=blue,
	filecolor=black,      
	urlcolor=red,
}
\usepackage{times}
\usepackage[paper=a4paper,left=25mm,right=25mm,top=20mm,bottom=25mm]{geometry}
\usepackage{graphicx}
\usepackage{epstopdf}
\usepackage{fancyhdr}
\usepackage{fancybox}
\usepackage[]{amsmath}
\usepackage{subfigure}
\usepackage{float}
\usepackage[normalem]{ulem} 
\usepackage{amssymb}
\usepackage{mathrsfs}
\usepackage{tabularx}
\usepackage{booktabs}
\usepackage{bm}
\usepackage{natbib}
\usepackage{amsthm}

\journal{XXX}

\begin{document}
\begin{frontmatter}
\title{Boundary sources of velocity gradient tensor and its invariants}

\author[1]{Tao Chen\corref{mycorrespondingauthor}}
\cortext[mycorrespondingauthor]{Corresponding author}
\ead{chentao2023@njust.edu.cn}
\author[2]{Jie-Zhi Wu}
\ead{jie\_zhi\_wu@163.com}
\author[3]{Tianshu Liu}
\ead{tianshu.liu@wmich.edu}
\author[4]{Jie Yao}

\address[1]{School of Physics, Nanjing University of Science and Technology, Nanjing 210094, China}
\address[2]{State Key Laboratory for Turbulence and Complex Systems, College of Engineering, Peking University, Beijing 100871, China}
\address[3]{Department of Mechanical and Aerospace Engineering, Western Michigan University, Kalamazoo, Michigan, 49008, USA}
\address[4]{Advanced Research Institute of Multidisciplinary Sciences, Beijing Institute of Technology, Beijing 100081, China}

%
%
\begin{abstract}
The present work elucidates the boundary behaviors of the velocity gradient tensor ($\bm{A}\equiv\bm{\nabla}\bm{u}$) and its principal invariants ($P,Q,R$) for compressible flow interacting with a stationary rigid wall. Firstly, it is found that the well-known Caswell formula exhibits an inherent physical structure being compatible with the normal-nilpotent decomposition, where both the strain-rate and rotation-rate tensors contain the physical effects from the spin component  of the vorticity. 
Secondly, we derive the kinematic and dynamic forms of the boundary $\bm{A}$-flux from which the known boundary fluxes can be recovered by applying the symmetric-antisymmetric decomposition.
Then, we obtain the explicit expression of the boundary $Q$ flux as a result of the competition among the boundary fluxes of squared dilatation, enstrophy and squared strain-rate.
Importantly, we emphasize that both the coupling between the spin and surface pressure gradient, and the spin-curvature quadratic interaction ($\bm{s}_{w}\bm{\cdot}\bm{K}\bm{\cdot}\bm{s}_{w}$), are \textit{not} responsible for the generation of the boundary $Q$ flux, although they contribute to both the boundary fluxes of enstrophy and squared strain-rate. 
Moreover, we prove that the boundary $R$ flux must vanish on a stationary rigid wall.
Finally, the boundary fluxes of the invariants of the strain-rate and rotation-rate tensors are also discussed. It is revealed that the boundary flux of the third invariant of the strain-rate tensor is proportional to the wall-normal derivative of the vortex stretching term ($\bm{\omega}\bm{\cdot}\bm{D}\bm{\cdot}\bm{\omega}$), which serves as a source term accounting for the the spatiotemporal evolution rate of the wall-normal enstrophy flux.
These theoretical results provide a unified description of boundary vorticity and vortex dynamics, which could be valuable in understanding the formation mechanisms of complex near-wall coherent structures and the boundary sources of flow noise.
\end{abstract}

\end{frontmatter}
\section{\textcolor{red}{Introduction}}
\subsection{\textcolor{red}{Velocity gradient tensor and its principal invariants}}\label{SS1new}
The velocity gradient tensor (VGT, $\bm{A}\equiv\bm{\nabla}\bm{u}$) has been widely applied to understand the fundamental flow physics and to predict the evolution of coherent structures in turbulence, which has shown great advantages in its potential to boost the 
the cross-communication among the branches of turbulence dynamics, statistics and coherent structures~\citep{Meneveau2011,Johnson2024ARFM}. The eigenvalues $\lambda_{i} (i=1,2,3)$ of the VGT are the solutions of the characteristic equation
\begin{eqnarray}\label{cc1}	det\left(\lambda\bm{I}-\bm{A}\right)=\lambda^{3}+P\lambda^{2}+Q\lambda+R=0,
\end{eqnarray}
where $P$, $Q$ and $R$ are the first, second and third principal invariants of $\bm{A}$, respectively defined by~\citep{Chong1990,ChuLu2013}
\begin{subequations}\label{EQ}
	\begin{equation}\label{EQ1}
		P\equiv-tr(\bm{A})=-\vartheta=-(\lambda_{1}+\lambda_{2}+\lambda_{3}),	
	\end{equation}
	\begin{equation}\label{EQ2}
		Q\equiv\frac{1}{2}\left(P^{2}-tr(\bm{A}^{2})\right)=\frac{1}{2}\left(\vartheta^{2}-tr(\bm{A}^{2})\right)=\lambda_{1}\lambda_{2}+\lambda_{2}\lambda_{3}+\lambda_{1}\lambda_{3},
	\end{equation}
	\begin{equation}\label{EQ3}
		R\equiv-det\left(\bm{A}\right)=-\lambda_{1}\lambda_{2}\lambda_{3}=\frac{1}{3}\left[\vartheta^{3}-3\vartheta Q-tr(\bm{A}^{3})\right],
	\end{equation}
\end{subequations}
where $\vartheta\equiv\bm{\nabla}\bm{\cdot}\bm{u}$ is the dilatation (namely, the velocity divergence). The last equality in~\eqref{EQ3} is obtained by applying the \textit{Cayley-Hamilton theorem}.

The discriminant of~\eqref{cc1} is given by
\begin{eqnarray}\label{discriminant}
	\Delta=\frac{27}{4}R^{2}+\left(P^{3}-\frac{9}{2}PQ\right)R+\left(Q^{3}-\frac{1}{4}P^{2}Q^{2}\right).
\end{eqnarray}
In the region $\Delta\leq{0}$, $\bm{A}$ has three real eigenvalues $\lambda_{i} (i=1,2,3)$; in the region $\Delta>{0}$, $\bm{A}$ has one real ($\lambda_{3}$) and two complex-conjugate ($\lambda_{1,2}=\lambda_{cr}\pm{i}\lambda_{ci}$) eigenvalues.
According to the sign of $\Delta$, the flow topology of complex flows can be classified in $(P,Q,R)$ space or $(Q,R)$ space (for a fixed $P$) using critical point terminology for both the incompressible and compressible flows~\citep{ChuLu2013,Meneveau2011}. In addition, starting from the Navier-Stokes (NS) equations, the Lagrangian evolution equations of the VGT and its invariants were also derived and interpreted in flows of various types~\citep{Johnson2024ARFM}, leading to the simplified descriptions of the restricted Euler dynamics~\citep{Vieillefosse1982,Vieillefosse1984,Cantwell1992}.

An interesting and important observation is that 
the existing studies of the VGT and its invariants~\citep{Meneveau2011,ChuLu2013,Johnson2024ARFM} are confined to the physical processes in
unbounded flows or the flow structures in the fluid interior. 
Since their Lagrangian evolution equations cannot be simply written in single-variable forms, their sources in the fluid interior are difficult to be determined~\citep{MaoFeng2022}, which thereby highlights the importance of specifying their boundary sources (with perfect certainty).

From a physical perspective, since the considered physical quantities cannot penetrate into the solid side of the boundary, their \textit{boundary fluxes} naturally become the \textit{boundary sources} which describe their local creation rates on the boundary. From a pure mathematical perspective, all these governing equations are one-order higher than the NS equations in terms of the highest-order derivatives. Once generalized to the wall-bounded flows, the boundary values and fluxes of the VGT and its invariants immediately become the definite conditions of the Dirichlet and Neumann types, respectively. 
To the best of the authors' knowledge, a general theory that reveals the physical structures of the boundary values and sources of the VGT and associated invariants (including their relationship with the established boundary vorticity and dilatation dynamics) is not available in the existing literature, which becomes the main objective of the present study. 

\subsection{\textcolor{red}{Kinematic decompositions of velocity gradient tensor}}\label{SS2new}
To acquire more detailed knowledge of flow structures, the kinematic decompositions of the VGT are performed based on different physical considerations.
The conventional decomposition of the VGT is the \textit{symmetric-antisymmetric decomposition} (SAD) or the \textit{Cauchy-Stokes decomposition}~\citep{WuJZ2015book}:
\begin{eqnarray}\label{SAD}
	\bm{A}=\bm{D}+\bm{\Omega},~\bm{D}\equiv\frac{1}{2}\left(\bm{A}+\bm{A}^{T}\right),~\bm{\Omega}\equiv\frac{1}{2}\left(\bm{A}-\bm{A}^{T}\right),
\end{eqnarray}
where $\bm{D}$ and $\bm{\Omega}$ are the strain-rate tensor and the rotation-rate tensor, respectively.
The latter has one-to-one correspondence to the vorticity  $\bm{\omega}\equiv\bm{\nabla}\times\bm{u}$ through the relation $\bm{\omega}=\bm{\epsilon}\bm{:}\bm{\Omega}$ and $\bm{\Omega}=\bm{\epsilon}\bm{\cdot}\bm{\omega}/2$, where $\bm{\epsilon}$ is the permutation tensor. 
Nevertheless, the inherent limitation of the SAD is that one can neither distinguish the deformation and shearing effects in the strain-rate tensor nor identify the orbital rotation and the spin in the rotation-rate tensor, as firstly noticed by~\citet{Kolar2004,Kolar2007IJHMF}.

Soon thereafter, applying the real Schur theorem to the VGT yields the \textit{normal-nilpotent decomposition} (NND)~\citep{LiZhen2014}:
\begin{eqnarray}\label{NND}
	\bm{A}=\bm{N}+\bm{S},
\end{eqnarray}
where $\bm{N}$ and $\bm{S}$ represent the normal and nilpotent tensors, respectively.
A key implication from such decomposition is that the vorticity $\bm{\omega}$ is naturally split as the sum of the \textit{orbital rotation} $\bm{\psi}$ and the \textit{spin} $\bm{s}$~\citep{LiuCQ2018,GaoLiu2019}:
  \begin{eqnarray}\label{omega_psi_s}
  	\bm{\omega}=2\bm{\psi}_{\Delta>0}+\bm{s},
  \end{eqnarray}
which lays a rational foundation for analyzing the evolution and interaction of axial vortex and shear layer. The orbital rotation $\bm{\psi}$ represents the angular velocity of a fluid element with respect to the curvature center of its pathline at a time instant. The spin $\bm{s}$ is solely determined by the nilpotent tensor $\bm{S}$, illuminating the pure shearing effect typically observed in a shear layer. 
 
A clear physical picture of vortex kinematics can be revealed by~\eqref{omega_psi_s}, being compatible with the observed forms of the dominated flow structures including the winding shear layer and the axial vortex. In the region with $\Delta\leq{0}$, the vorticity $\bm{\omega}$ is completely determined by the spin $\bm{s}$ without the orbital rotation ($\bm{\psi}=\bm{0}$). In contrast, in the region with $\Delta>0$, a shear layer rolls up to form an axial vortex, which displays both the orbital rotation $\bm{\psi}$ and the spin $\bm{s}$. For the potential flow ($\bm{\omega}=\bm{0}$), one still has $2\bm{\psi}+\bm{s}=\bm{0}$ in the region with $\Delta>0$, which indicates a mutual cancellation mechanism between the orbital rotation and the spin.
The applications of this new decomposition in the analysis of turbulence physics have been reported in limited studies, e.g.,~\citet{Keylock2018} and~\citet{JiangXY2022}.

\subsection{\textcolor{red}{Caswell formula and normal-nilpotent decomposition}}\label{SS3new}
The VGT can be decomposed as $\bm{A}=\vartheta\bm{I}+2\bm{\Omega}-\bm{B}$, where $\bm{B}\equiv\vartheta\bm{I}-\bm{A}^{T}$ is the surface deformation tensor which determines the relative rate of change of the small material surface element $\delta\bm{\Sigma}=\delta{\Sigma}\bm{n}$: $\delta{\Sigma}^{-1}D_{t}\delta\bm{\Sigma}=\bm{n}\bm{\cdot}\bm{B}=-(\bm{n}\times\bm{\nabla})\times\bm{u}$~\citep{WuJZ2005JFM}, fully determined by the velocity over and geometry of the surface [$D_{t}\equiv{D/Dt}$ is the material derivative]. On a stationary rigid wall, we find that
the vorticity is equal to the spin ($\bm{\omega}_{w}=\bm{s}_{w}$), extending along the tangential direction of the boundary. Therefore, the tangential-normal decomposition of $\bm{A}_{w}$ is given by
\begin{eqnarray}\label{A_w}
	\bm{A}_{w}={\vartheta}_{w}\bm{nn}+\bm{n}\left(\bm{s}_{w}\times\bm{n}\right)={\vartheta}_{w}\bm{nn}+\bm{n}\bm{\xi},
\end{eqnarray}
where $\bm{\xi}\equiv\bm{s}_{w}\times\bm{n}$ is a vector perpendicular to the spin and $\bm{n}$ is the unit normal vector directed into the fluid. $\bm{\tau}\equiv\mu\bm{\xi}$ is the skin friction vector ($\mu$ is the dynamic viscosity), which represents the wall shear stress exerted by the ambient viscous fluid. Then, a special on-wall orthonormal frame $(\bm{e}_{1},\bm{e}_{2},\bm{e}_{3})$ [referred to as the $\bm{\tau}$-frame in~\citet{WuJZ2000POF}] is naturally introduced as $\bm{e}_{1}\equiv\bm{\xi}/\lVert\bm{\xi}\rVert$, $\bm{e}_{2}\equiv\bm{s}_{w}/\lVert\bm{s}_{w}\rVert$ and $\bm{e}_{3}\equiv\bm{n}$.

Interestingly, we find that~\eqref{A_w} automatically implies a NND structure:
\begin{eqnarray}\label{AwNND}
	\bm{A}_{w}=\bm{N}_{w}+\bm{S}_{w},~\bm{N}_{w}=\vartheta_{w}\bm{n}\bm{n},
	~\bm{S}_{w}=\bm{n}\bm{\xi}.
\end{eqnarray}
By~\eqref{SAD} and~\eqref{AwNND}, the on-wall values of $\bm{D}$ and $\bm{\Omega}$ are respectively expressed as
\begin{eqnarray}\label{Caswell}
\bm{D}_{w}=\vartheta_{w}\bm{n}\bm{n}+\frac{1}{2}\left(\bm{n}\bm{\xi}+\bm{\xi}\bm{n}\right),~
\bm{\Omega}_{w}=\frac{1}{2}\left(\bm{n}\bm{\xi}-\bm{\xi}\bm{n}\right),
\end{eqnarray}
where the first equality is just the well-known Caswell formula on a stationary wall~\citep{Caswell1967}. 
On one hand, \textit{both $\bm{D}_{w}$ and $\bm{\Omega}_{w}$ contain the spin effects}, thereby challenging the widely accepted perspective that $\bm{D}_{w}$ represents the pure deformation of a fluid element. In this sense, the Caswell formula had anticipated the necessity of the NND prior to its formal introduction. On the other hand, the whole boundary vorticity dynamics is fully governed by the spin field whose sources and sinks determine the creation of the wall-normal vorticity component near the wall, satisfying $[\partial_{n}\omega_{n}]_{w}=-\bm{\nabla}_{\pi}\bm{\cdot}\bm{s}_{w}=\bm{n}\bm{\cdot}(\bm{\nabla}_{\pi}\times\bm{\xi})$ ($\bm{\nabla}_{\pi}$ is the surface tangential gradient operator).

It readily follows from~\eqref{EQ} and~\eqref{AwNND} that the boundary values of the invariants and the discriminant are
\begin{eqnarray}\label{m5}
	P_{w}=-\vartheta_{w},~Q_{w}=R_{w}=\Delta_{w}=0.
\end{eqnarray}
 
\subsection{\textcolor{red}{Known boundary fluxes}}\label{SS4new}
Without losing generality, the linear diffusion approximation~\citep{WuJZ2015book} is adopted in the following discussion so that all the viscosity coefficients are treated as constants.

\subsubsection{\textcolor{red}{Boundary vorticity and enstrophy fluxes}}\label{BVF_BEF_review}
The vorticity equation can be obtained from the anti-symmetric part of the evolution equation of $\bm{A}$:
\begin{eqnarray}\label{vorticity_equation}
	D_{t}\bm{\omega}+\bm{B}\bm{\cdot}\bm{\omega}=\bm{\nabla}T\times\bm{\nabla}s+\nu\nabla^{2}\bm{\omega},
\end{eqnarray}
where $T$ is the temperature, $s$ is the specific entropy; $\nu=\mu/\rho$ is the kinematic viscosity ($\rho$ is the density).
It is noticed that~\eqref{vorticity_equation} is one-order higher than the NS equations in terms of the diffusion term. Therefore, in addition to the velocity adherence boundary condition,
one has to add a new boundary condition to avoid possible spurious solutions caused by 
raising the order. The most natural choice is the acceleration adherence condition by which restricting the 
NS equations to the wall yields the boundary vorticity flux (BVF)~\citep{Lighthill1963,Panton1984,WuJZWuJM1998}. The BVF on a stationary rigid wall is expressed as
\begin{subequations}\label{BVF_abc}
	\begin{eqnarray}\label{total_BVF}
		\bm{\sigma}\equiv\nu[\partial_{n}\bm{\omega}]_{w}=\bm{\sigma}_{p}+\bm{\sigma}_{vis},
	\end{eqnarray}
	\begin{eqnarray}\label{BVF_p}
		\bm{\sigma}_{p}\equiv\rho^{-1}\bm{n}\times\bm{\nabla}_{\pi}\Pi_{w},~\Pi\equiv{p}-\mu_{\vartheta}\vartheta,
	\end{eqnarray}
	\begin{eqnarray}\label{BVF_vis}
		\bm{\sigma}_{vis}\equiv\nu\bm{K}\bm{\cdot}\bm{s}_{w}-\nu\left(\bm{\nabla}_{\pi}\bm{\cdot}\bm{s}_{w}\right)\bm{n},
	\end{eqnarray}
\end{subequations}
where $p$ is the pressure, $\bm{K}\equiv-\bm{\nabla}_{\pi}\bm{n}$ is the surface curvature tensor, $\mu_{\vartheta}\equiv\mu_{b}+4\mu/3$ is the longitudinal viscosity with $\mu_{b}$ being the bulk viscosity.
Equation~\eqref{BVF_p} reveals a local dynamic causal mechanism of the boundary vorticity creation by the surface pressure gradient (with a dilatational correction for pressure), under the constraint of the no-slip boundary condition. Equation~\eqref{BVF_vis} represents the pure 3D viscous contributions to the total BVF, which are
caused by the coupling between surface curvature and boundary spin, as well as the sources/sinks
of the boundary spin field. A complete analysis of the BVF with the aid of numerical simulations can be found in~\citet{ChenYu2017}.

The boundary enstrophy flux (BEF) $F_{\Omega}\equiv\nu[\partial_{n}\Omega]_{w}$ ($\Omega\equiv\omega^2/2$ is the enstrophy) is directly related to the BVF through the relation $F_{\Omega}=\bm{\sigma}\bm{\cdot}\bm{s}_{w}$~\citep{WuJZ1995,Liu2016MST,ChenTao2019POF}.
By taking a dot product of $\bm{s}_{w}$ with~\eqref{BVF_abc}, one can obtain the explicit expression of the BEF as
\begin{eqnarray}\label{BEF}
	F_{\Omega}
	=\rho^{-1}\bm{\xi}\bm{\cdot}\bm{\nabla}_{\pi}\Pi_{w}
	+\nu\bm{s}_{w}\bm{\cdot}\bm{K}\bm{\cdot}\bm{s}_{w}.
\end{eqnarray}
In the right hand side of~\eqref{BEF}, the first term represents the interaction between the transverse and longitudinal fields while the second term denotes the quadratic coupling between the boundary spin and the surface curvature tensor. Recently, the BEF has found a new application in studying the surface flow structures:
the global skin friction field can be inferred from the surface pressure field measured by
the pressure sensitive paint (PSP) by modeling the BEF properly~\citep{Liu2016MST,ChenTao2019POF}.

\subsubsection{\textcolor{red}{Boundary dilatation flux}}\label{BDF_review}
The dilatation equation can be derived by evaluating the trace of the evolution equation of $\bm{A}$ and then eliminating the specific enthalpy.
For the \textit{isentropic} flow, the dilatation satisfies the following advective-wave equation~\citep{MaoFeng2022}:
\begin{eqnarray}\label{AW_theta}
	D_t^2\vartheta+D_{t}\bm{u}\bm{\cdot}\bm{\nabla}\vartheta
	-\nabla^{2}\left(c^2\vartheta\right)=Q_{0},
\end{eqnarray}
where the source term is $Q_{0}\equiv-2D_{t}tr(\bm{A}^2)-2tr(\bm{A}^{3})+D_{t}\bm{u}\bm{\cdot}\left(\bm{\nabla}\times\bm{\omega}\right)$, $c^{2}={\gamma RT}$ is the square of the local sound speed, $\gamma$ is the specific heat ratio and $R$ is the gas constant.

The boundary longitudinal dynamics, caused by the on-wall viscous coupling among the transverse and longitudinal processes, is usually much more complicated than the boundary transverse dynamics~\citep{WuJZ2015book,MaoFeng2022b}.
Similar to the BVF, the boundary dilatation flux (BDF) $F_{\vartheta}\equiv[\partial_{n}\vartheta]_{n}$ is introduced and used as a required boundary condition to solve~\eqref{AW_theta}. 
Physically, the BDF does not exist on the other side of the wall, which implies a creation mechanism of dilatation at the boundary.
By performing the singular perturbation analysis, a simplified expression of the BDF is obtained as~\citep{MaoFeng2022b}:
\begin{eqnarray}\label{BDF}
	F_{\vartheta}=\frac{\gamma\nu}{c^2}(\partial_{t}+\vartheta_{w})\left(\bm{n}\times\bm{\nabla}_{\pi}\right)\bm{\cdot}\bm{s}_{w}
	+\frac{\gamma\nu}{c^2}\left(\bm{s}_{w}\bm{\cdot}\bm{K}\bm{\cdot}\bm{s}_{w}-[\partial_{n}\Omega]_{n}\right)
	-\frac{\gamma\nu_{\theta}}{c^2}\bm{\xi}\bm{\cdot}\bm{\nabla}_{\pi}\vartheta_{w}.
\end{eqnarray}
By using~\eqref{BEF}, $c^2=\gamma{p}/\rho$ and $\left(\bm{n}\times\bm{\nabla}_{\pi}\right)\bm{\cdot}\bm{s}_{w}=\bm{\nabla}_{\pi}\bm{\cdot}\bm{\xi}$,~\eqref{BDF} can further be written as
\begin{eqnarray}\label{BDF2}
	F_{\vartheta}&=&\frac{\gamma\nu}{c^2}(\partial_{t}+\vartheta_{w})\left[\bm{\nabla}_{\pi}\bm{\cdot}\bm{\xi}\right]-\bm{\xi}\bm{\cdot}\bm{\nabla}_{\pi}\ln{p}_{w}\nonumber\\
	&=&-\bm{\xi}\bm{\cdot}\bm{\nabla}_{\pi}\ln{p}_{w}+\mathcal{O}(Re^{-1/2}).
\end{eqnarray}
It is emphasized that, the first term in the right hand side of~\eqref{BDF2} is generally of the order $\mathcal{O}(Re^{-1/2})$.
Therefore, the BDF is dominated by the coupling between the transverse field ($\bm{\xi}$ or $\bm{s}_{w}$) and the surface pressure gradient with the order of $O(Re^{1/2})$, which also generates the BEF.

\section{\textcolor{red}{Boundary flux of velocity gradient tensor}}\label{BFVGT}
The boundary flux of the VGT (namely, the \textit{boundary $\bm{A}$-flux} for short) is defined as the wall-normal derivative of $\bm{A}$ at the wall.
By using the velocity adherence condition ($\bm{u}_{w}=\bm{0}$), \textit{the kinematic form of the boundary $\bm{A}$-flux} is derived as (\ref{New_App_1})
\begin{eqnarray}\label{hh8}
\left[\partial_{n}\bm{A}\right]_{w}
&=&K\bm{A}_{w}+(\bm{\nabla}_{\pi}\vartheta_{w})\bm{n}-\vartheta_{w}\bm{K}+\bm{\nabla}_{\pi}\bm{\xi}\nonumber\\
& &+\bm{n}\bm{n}(\left[\partial_{n}\vartheta\right]_{w}-\bm{\nabla}_{\pi}\bm{\cdot}\bm{\xi})+\bm{n}\left\{\bm{\nabla}_{\pi}\vartheta_{w}+([\partial_{n}\bm{\omega}]_{w}-\bm{K}\bm{\cdot}\bm{s}_{w})\times\bm{n}\right\},
\end{eqnarray}
where $K\equiv-\bm{\nabla}_{\pi}\bm{\cdot}\bm{n}=tr(\bm{K})$ is twice the mean curvature of the boundary surface. Equation~\eqref{hh8} is the core of the present study, which will be employed to derive other results in the following text.
Although the derivation is apparently concerned with only the fluid kinematics, viscosity must be involved to enforce the no-slip boundary condition, which is compatible with the viscid nature of the NS equations.

It is emphasized that both $[\partial_{n}\bm{\omega}]_{w}$ and $[\partial_{n}\vartheta]_{w}$ are embodied in the kinematic form of the boundary $\bm{A}$-flux, which can be replaced by using~\eqref{BVF_abc} ,~\eqref{BDF} and~\eqref{BDF2} to associate them with the fundamental surface quantities. For neatness, $[\partial_{n}\vartheta]_{w}$ is not expanded in the following discussion. It follows from~\eqref{BVF_abc} that $\left([\partial_{n}\bm{\omega}]_{w}-\bm{K}\bm{\cdot}\bm{s}_{w}\right)\times\bm{n}=\nu^{-1}\bm{\nabla}_{\pi}H_w$ ($H\equiv\Pi/\rho=p/\rho-\nu_{\vartheta}\vartheta$) by which \textit{the dynamic form of the boundary $\bm{A}$-flux} can be derived from~\eqref{hh8}:
\begin{eqnarray}\label{hh9}
[\partial_{n}\bm{A}]_{w}&=&\bm{n}\bm{n}([\partial_{n}\vartheta]_{w}+K\vartheta_{w}-\bm{\nabla}_{\pi}\bm{\cdot}\bm{\xi})
+\bm{n}\left(K\bm{\xi}+\nu^{-1}\bm{\nabla}_{\pi}H_{w}\right)\nonumber\\
& &+(\bm{\nabla}_{\pi}\vartheta_{w})\bm{n}+\bm{n}(\bm{\nabla}_{\pi}\vartheta_{w})-\vartheta_{w}\bm{K}+\bm{\nabla}_{\pi}\bm{\xi}.
\end{eqnarray}
Here, $[\partial_{n}\vartheta]_{w}$ [\eqref{BDF} and~\eqref{BDF2}] and $\nu^{-1}\bm{\nabla}_{\pi}H_{w}$ are the direct results of on-wall dynamics while the remaining terms are originated from pure kinematics. 
The terms $K\vartheta_{w}\bm{nn}$ and $K\bm{n}\bm{\xi}$ come from the kinematic decomposition of $K\bm{A}_{w}$.
 In addition, there exist three terms involving $\bm{\xi}$ [of the order $\mathcal{O}(Re^{1/2})$] with two of them representing its tangential divergence and gradient, respectively. These terms highlight the crucial roles of spin in the near-wall fluid dynamics.
 All the remaining terms associated with the boundary dilatation are generally of the order $\mathcal{O}(1)$, which physically represent the longitudinal waves as the by-products of near-wall flow.

At the high-Reynolds-number limit ($\nu\rightarrow{0}$), $\nu^{-1}\bm{\nabla}_{\pi}H_{w}=\mathcal{O}(Re)$ is the primary leading term and
\begin{eqnarray}\label{Limit_A}
	\lim_{\nu\rightarrow{0}}\nu\left[\partial_{n}\bm{A}\right]_{w}=\frac{1}{\rho_w}\bm{n}\bm{\nabla}_{\pi}p_{w}=\mathcal{O}(1),
\end{eqnarray}
which holds inside the material vortex sheet attached to the boundary with its thickness $\delta\rightarrow{0}$.
In other words, the surface pressure gradient becomes the dominant boundary sources for the boundary $\bm{A}$-flux, where the viscosity guarantees the created diffusive flux to be of $\mathcal{O}(1)$ as $Re\rightarrow\infty$.

The symmetric and anti-symmetric parts of~\eqref{hh9} yield twice of the boundary fluxes of $\bm{D}$ and $\bm{\Omega}$, respectively expressed as
\begin{eqnarray}\label{BSRF}
2\left[\partial_{n}\bm{D}\right]_{w}&=&2(\bm{\nabla}_{\pi}\vartheta_{w})\bm{n}+2\bm{n}(\bm{\nabla}_{\pi}\vartheta_{w})-2\bm{K}\vartheta_{w}\nonumber\\
& &+\bm{n}\left(K\bm{\xi}+\nu^{-1}\bm{\nabla}_{\pi}H_{w}\right)+\left(K\bm{\xi}+\nu^{-1}\bm{\nabla}_{\pi}H_{w}\right)\bm{n}\nonumber\\
& &+\bm{\nabla}_{\pi}\bm{\xi}
+\bm{\nabla}_{\pi}\bm{\xi}^{T}\nonumber\\
& &+2\bm{nn}\left\{[\partial_{n}\vartheta]_{w}+K\vartheta_{w}-(\bm{\nabla}_{\pi}\bm{\cdot}\bm{\xi})\right\},
\end{eqnarray}
\begin{eqnarray}\label{BOmegaF}
	2\left[\partial_{n}\bm{\Omega}\right]_{w}&=&\bm{n}\left(K\bm{\xi}+\nu^{-1}\bm{\nabla}_{\pi}H_{w}\right)-\left(K\bm{\xi}+\nu^{-1}\bm{\nabla}_{\pi}H_{w}\right)\bm{n}\nonumber\\
	& &+\bm{\nabla}_{\pi}\bm{\xi}
	-\bm{\nabla}_{\pi}\bm{\xi}^{T}.
\end{eqnarray}
It is noted that the trace of~\eqref{BSRF} recovers the BDF in~\eqref{BDF2}. The dual vector form of~\eqref{BOmegaF} gives the BVF in~\eqref{BVF_abc}, where the detailed mathematical proof is documented in~\ref{recovery}.
By now, we have established an intrinsic theory of the boundary $\bm{A}$-flux which indeed recovers the known BDF and BVF by applying the SAD. 

\section{\textcolor{red}{Boundary fluxes of $Q$ and $R$}}\label{Theory_BQF}
For the ideal perfect gas undergoing an isentropic process, the fluctuating density wave satisfies the classical Phillips equation~\citep{Phillips1960JFM}:
\begin{eqnarray}\label{Phillips equation}
D_{t}^{2}\mathcal{R}-\bm{\nabla}\bm{\cdot}\left(c^2\bm{\nabla}\mathcal{R}\right)=tr(\bm{A}^{2}),~\mathcal{R}\equiv\ln\left(\frac{\rho}{\rho_{0}}\right),
\end{eqnarray}
where $\mathcal{R}$ describes the relative strength of the density disturbance ($\rho^{\prime}\equiv\rho-\rho_{0}$), $\rho_{0}$ is the reference density. For a small disturbance ($\lvert\rho^{\prime}\rvert\ll{\rho_{0}}$), we have $\mathcal{R}=\ln(1+\rho^{\prime}/\rho_{0})\approx\rho^{\prime}/\rho_{0}$.  Because of $tr(\bm{A}^2)=\vartheta^{2}-2Q$, $\vartheta^{2}=(D_t\mathcal{R})^{2}$ can be moved to the left hand side of~\eqref{Phillips equation} as a low-order nonlinear term:~\citep{MaoFeng2022}
\begin{eqnarray}\label{Phillips equation2}
	D_{t}^{2}\mathcal{R}-(D_{t}\mathcal{R})^{2}-\bm{\nabla}\bm{\cdot}\left(c^2\bm{\nabla}\mathcal{R}\right)=-2Q,
\end{eqnarray}
with $-2Q$ being a more concentrated kinematic source of sound than $tr(\bm{A}^{2})$. Physically, $Q$ represents a pseudo inviscid work rate done by the surface deformation over the boundary of a fluid element~\citep{MaoFeng2020JFM}.
For incompressible flow, the pressure satisfies the Poisson equation $\nabla^{2}(p/\rho)=2Q$, which can be written as a weighted integral of $Q$ over the whole space~\citep{Johnson2024ARFM}.
However, to the best of the authors' knowledge, the boundary $Q$ flux, describing the creation mechanism of $Q$ on the boundary, was never investigated from a theoretical perspective.

By using~\eqref{EQ2} and~\eqref{BQF_T2}, the boundary $Q$ flux is evaluated as
\begin{eqnarray}\label{BQFW}
\left[\partial_{n}Q\right]_{w}&=&\vartheta_{w}[\partial_{n}\vartheta]_{w}
-\frac{1}{2}[\partial_{n}tr(\bm{A}^2)]_{w}\nonumber\\
&=&(\bm{\nabla}_{\pi}\bm{\cdot}\bm{\xi})\vartheta_{w}
-\bm{\xi}\bm{\cdot}\bm{\nabla}_{\pi}\vartheta_{w}
-K\vartheta_{w}^{2}
-\bm{\xi}\bm{\cdot}\bm{K}\bm{\cdot}\bm{\xi}.
\end{eqnarray}
Equation~\eqref{BQFW} shows that the boundary $Q$ flux can be generated by the coupling between the longitudinal and transverse fields, and their interplay with the surface curvature. 
The divergence ($\bm{\nabla}_{\pi}\bm{\cdot}\bm{\xi}$) is an intriguing surface physical quantity, which is closely associated with the strong sweep and ejection events (intermittency) in the viscous sublayer of near-wall turbulence~\citep{ChenTao2021POF,Chen2023PhysicaD}. It is found that the last quadratic term satisfies an exact identity: $	\bm{s}_{w}\bm{\cdot}\bm{K}\bm{\cdot}\bm{s}_{w}+\bm{\xi}\bm{\cdot}\bm{K}\bm{\cdot}\bm{\xi}=2K\Omega_{w}$,
where the quadratic term $\bm{s}_{w}\bm{\cdot}\bm{K}\bm{\cdot}\bm{s}_{w}$ contributes to the BEF in~\eqref{BEF}.

From~\eqref{EQ2} and~\eqref{SAD}, $Q$ could be alternatively written as
$2Q=\vartheta^{2}+\Omega-S$,
where $\Omega\equiv\omega^2/2=-tr(\bm{\Omega}^2)=\bm{\Omega}\bm{:}\bm{\Omega}$ is the enstrophy and $S\equiv tr(\bm{D}^2)=\bm{D}\bm{:}\bm{D}$ is the squared strain-rate. In other words, $Q$ represents the balance among squared dilatation, enstrophy and squared strain-rate.
Therefore, we have
\begin{eqnarray}\label{c0}
2[\partial_{n}Q]_{w}
=2\vartheta_{w}[\partial_{n}\vartheta]_{w}+[\partial_{n}\Omega]_{w}-[\partial_{n}S]_{w}.
\end{eqnarray}
We notice that both $\Omega$ and $S$ can be expressed in terms of $\bm{A}$:
\begin{eqnarray}\label{trace_decomp}
\Omega=\frac{1}{2}\left[tr(\bm{A}\bm{\cdot}\bm{A}^{T})-tr(\bm{A}^2)\right],~
	{S}=\frac{1}{2}\left[tr(\bm{A}\bm{\cdot}\bm{A}^{T})+tr(\bm{A}^2)\right].
\end{eqnarray}	
The boundary fluxes of the two traces in~\eqref{trace_decomp} are evaluated in~\ref{BFTrace} by which we obtain
\begin{subequations}\label{ab45}
\begin{eqnarray}\label{a45}
	[\partial_{n}\Omega]_{w}=\nu^{-1}\bm{\xi}\bm{\cdot}\bm{\nabla}_{\pi}H_{w}
	+\bm{s}_{w}\bm{\cdot}\bm{K}\bm{\cdot}\bm{s}_{w},
\end{eqnarray}
\begin{eqnarray}\label{b45}
[\partial_{n}S]_{w}&=&2\vartheta_{w}[\partial_{n}\vartheta]_{w}+2K\vartheta_{w}^{2}\nonumber\\
& &+2\left[\bm{\xi}\bm{\cdot}\bm{\nabla}_{\pi}\vartheta_{w}-(\bm{\nabla}_{\pi}\bm{\cdot}\bm{\xi})\vartheta_{w}\right]+\nu^{-1}\bm{\xi}\bm{\cdot}\bm{\nabla}_{\pi}H_{w}\nonumber\\
& &+\bm{s}_{w}\bm{\cdot}\bm{K}\bm{\cdot}\bm{s}_{w}+2\bm{\xi}\bm{\cdot}\bm{K}\bm{\cdot}\bm{\xi}.
\end{eqnarray}
\end{subequations}
When multiplied by the kinematic viscosity $\nu$,~\eqref{a45} just gives the BEF in~\eqref{BEF} while~\eqref{b45} provides the intrinsic decomposition for the boundary squared-strain-rate flux (BSF). In the right hand side of~\eqref{b45}, the first two terms represent the boundary flux of the squared dilatation and its coupling with the surface mean curvature. The third and fourth terms describe the interaction among the transverse and longitudinal fields, featured by $\bm{\xi}$, $p_{w}$, $\vartheta_{w}$ and their tangential derivatives along the boundary.
The last two terms result from the interaction between the boundary vorticity and the surface curvature, being manifested as the two quadratic forms. Substituting~\eqref{a45} and~\eqref{b45} into~\eqref{c0} again yields~\eqref{BQFW}.

For incompressible flow interacting with a stationary wall,~\eqref{BQFW},~\eqref{a45} and~\eqref{b45} reduce to
\begin{equation}\label{BQFw1}
	[\partial_{n}Q]_{w}=
	-\bm{\xi}\bm{\cdot}\bm{K}\bm{\cdot}\bm{\xi},
\end{equation}
\begin{subequations}\label{BESFw1}
	\begin{equation}\label{BEFw1}
		[\partial_{n}\Omega]_{w}=\mu^{-1}\bm{\xi}\bm{\cdot}\bm{\nabla}_{\pi}p_{w}
		+\bm{s}_{w}\bm{\cdot}\bm{K}\bm{\cdot}\bm{s}_{w},
	\end{equation}
	\begin{equation}\label{BSFw1}
		[\partial_{n}S]_{w}=\mu^{-1}\bm{\xi}\bm{\cdot}\bm{\nabla}_{\pi}p_{w}
		+\bm{s}_{w}\bm{\cdot}\bm{K}\bm{\cdot}\bm{s}_{w}
		+2\bm{\xi}\bm{\cdot}\bm{K}\bm{\cdot}\bm{\xi}.
	\end{equation}
\end{subequations}
It is emphasized that both the BEF and BSF can be created by both the longitudinal-transverse and the transverse-geometric coupling mechanisms. The longitudinal-transverse coupling is determined by the interaction between the surface vorticity (or $\bm{\xi}$) and the surface pressure gradient. The transverse-geometric coupling involves only the diagonal components of the surface curvature tensor under the ${\bm{\tau}}$-frame. The cancellation between the common terms (including $\mu^{-1}\bm{\xi}\bm{\cdot}\bm{\nabla}_{\pi}p_{w}$ and $\bm{s}_{w}\bm{\cdot}\bm{K}\bm{\cdot}\bm{s}_{w}$) in~\eqref{BEFw1} and~\eqref{BSFw1} leads to the fact that the boundary $Q$ flux in~\eqref{BQFw1} that is solely determined by $-\bm{\xi}\bm{\cdot}\bm{K}\bm{\cdot}\bm{\xi}$.
In particular, if the wall is flat, the three scalar fluxes become quite simple:
\begin{eqnarray}\label{eq53}
[\partial_{n}Q]_{w}=0,~[\partial_{n}\Omega]_{w}=[\partial_{n}S]_{w}=\mu^{-1}\bm{\xi}\bm{\cdot}\bm{\nabla}_{\pi}p_{w}.
\end{eqnarray}
Equation~\eqref{eq53} implies that the boundary $Q$ flux cannot be created on a stationary flat wall. The unique boundary coupling mechanism that can generate the BEF or BSF is the interaction between $\bm{\xi}$ and $p_{w}$.

On a stationary wall, it is direct to show that $\bm{A}_{w}^{2}=\vartheta_{w}\bm{A}_{w}$,
which implies that
\begin{eqnarray}\label{m2}
[\partial_{n}tr(\bm{A}^3)]_{w}&=&3\vartheta_{w}\bm{A}_{w}^{T}\bm{:}[\partial_{n}\bm{A}]_{w}=\frac{3}{2}\vartheta_{w}\left[\partial_{n}tr(\bm{A}^2)\right]_w\nonumber\\
&=&3\vartheta_{w}^{2}[\partial_{n}\vartheta]_{w}-3\vartheta_{w}[\partial_{n}Q]_{w}.
\end{eqnarray}
Then, combining~\eqref{EQ3},~\eqref{m5} and ~\eqref{m2} yields
\begin{eqnarray}\label{BRF}
[\partial_{n}{R}]_{w}&=&\frac{1}{3}\left[\partial_{n}(\vartheta^3-3\vartheta Q)\right]_{w}-\vartheta_{w}^2\left[\partial_{n}\vartheta\right]_{w}
+\vartheta_{w}\left[\partial_{n}Q\right]_{w}\nonumber\\
&=&-\left[\partial_{n}\vartheta\right]_{w}Q_{w}=0.
\end{eqnarray}
Although the BDF $\left[\partial_{n}\vartheta\right]_{w}$ is generally non-zero, $Q_{w}=0$ (on a stationary wall) guarantees that the boundary $R$ flux must be zero. Therefore, we conclude that the boundary $R$ flux could only be regulated in the presence of non-stationary boundaries, for example, a moving and continuously deforming boundary.

\section{\textcolor{red}{Boundary fluxes of principal invariants of strain-rate and rotation-rate tensors}}\label{kkkkk}
\subsection{\textcolor{red}{General discussion}}
Like the VGT, $\bm{D}$ and $\bm{\Omega}$ also have their own sets of principal invariants represented by $(P_{D},Q_{D},R_{D})$ and $(P_{\Omega},Q_{\Omega},R_{\Omega})$, which are explicitly written as follows:
\begin{subequations}\label{o123}
\begin{eqnarray}\label{o1}
P_{D}=-\vartheta=-tr(\bm{D}),~P_{\Omega}=0,~P=P_{D},
\end{eqnarray}
\begin{eqnarray}\label{o2}
Q_{D}=\frac{1}{2}\left(\vartheta^{2}-tr(\bm{D}^{2})\right),
~Q_{\Omega}=-\frac{1}{2}tr(\bm{\Omega}^2),~Q=Q_{D}+Q_{\Omega},
\end{eqnarray}
\begin{eqnarray}\label{o3}
R_{D}=\frac{1}{3}\left(\vartheta^3-3\vartheta{Q}_{D}-tr(\bm{D}^3)\right),~R_{\Omega}=0,~R=R_{D}-\frac{1}{4}\bm{\omega}\bm{\cdot}\bm{D}\bm{\cdot}\bm{\omega}.
\end{eqnarray}
\end{subequations}

A proof of the last equality in~\eqref{o3} is given in~\ref{AppendixA}.
Obviously, according to~\eqref{o1}, both the boundary $P_{D}$ and $P$ fluxes are equal to the negative BDF [\eqref{BDF} and~\eqref{BDF2}].
In addition, it follows from~\eqref{o2} that
\begin{eqnarray}
2\left[\partial_{n}Q_{D}\right]_{w}=\left[\partial_{n}\vartheta^2\right]_{w}-\left[\partial_{n}S\right]_{w},~2\left[\partial_{n}Q_{\Omega}\right]_{w}=\left[\partial_{n}\Omega\right]_{w}.
\end{eqnarray}
The boundary $Q_{D}$ flux is determined by the boundary fluxes of squared dilatation and squared strain-rate [\eqref{b45}] while the boundary $Q_{\Omega}$ flux is equivalent to the BEF [\eqref{BEF} or~\eqref{a45}]. It is worth mentioning that the boundary flux of the squared-dilatation is balanced by the first term in the right hand side of~\eqref{b45}, resulting in zero net contribution to the boundary $Q_{D}$ flux.

Since the boundary $R$ flux is zero on a stationary wall [\eqref{BRF}], the last equality in~\eqref{o3} implies that
the boundary $R_{D}$ flux is proportional to the wall-normal gradient of the vortex stretching term in the enstrophy transport equation (\ref{AppendixB}):
\begin{eqnarray}\label{BRDF}
[\partial_{n}{R}_{D}]_{w}
&=&\frac{1}{4}\left[\partial_{n}(\bm{\omega}\bm{\cdot}\bm{D}\bm{\cdot}\bm{\omega})\right]_{w}\nonumber\\
&=&-\frac{1}{4}\left[\bm{\xi}\bm{\cdot}\bm{\nabla}_{\pi}\Omega_{w}
-2(\bm{\nabla}_{\pi}\bm{\cdot}\bm{\xi})\Omega_{w}
+\vartheta_{w}(\bm{s}_{w}\bm{\cdot}\bm{K}\bm{\cdot}\bm{s}_{w})\right],
\end{eqnarray}
In the right hand side of~\eqref{BRDF}, the first term represents the variation of the boundary enstrophy along a $\bm{\xi}$-line (or a skin friction line). The second term could be significant near the attachment and separation lines with high-magnitude skin friction divergence and surface enstrophy.
The last term implies a triple coupling mechanism among dilatation, vorticity and surface curvature tensor on the boundary. 
\subsection{\textcolor{red}{Role of boundary $R_{D}$ flux and further discussion}}
For flow past a stationary rigid wall, denoting the spatiotemporal evolution operator by $\mathcal{L}\equiv\partial_{t}-\nu\nabla^{2}$, the enstrophy transport equation can be written as
\begin{eqnarray}\label{enstrophy_equation}
\mathcal{L}\Omega=-\bm{u}\bm{\cdot}\bm{\nabla}\Omega-2\vartheta\Omega
+\bm{\omega}\bm{\cdot}\bm{D}\bm{\cdot}\bm{\omega}
+\bm{\omega}\bm{\cdot}\bm{\nabla}T\times\bm{\nabla}s
-\nu\bm{\nabla}\bm{\omega}\bm{:}\bm{\nabla}\bm{\omega}.
\end{eqnarray}
where in the right hand side, the first term is the enstrophy convection term; the second term represents the interaction between dilatation and enstrophy; the third term describes the effect of vortex stretching; the fourth term represents the baroclinic enstrophy generation as a result of the misalignment between the temperature and entropy gradients; the final term describes the dissipation of enstrophy.

By acting the wall-normal derivative operator $\partial_{n}$ on both sides of~\eqref{enstrophy_equation} and applying the result on the wall (mildly curved such that $\partial_{n}\mathcal{L}\approx\mathcal{L}\partial_{n}$), we obtain the spatiotemporal evolution rate of the wall-normal enstrophy flux at the wall:
\begin{eqnarray}\label{ss0}
\left[\mathcal{L}\partial_{n}\Omega\right]_{w}
=\Sigma_{1}+\Sigma_{2}+\Sigma_{3}+\Sigma_{4}+\Sigma_{5},
\end{eqnarray}
where the source terms can be expressed by using the surface physical quantities:
\begin{subequations}\label{five_source_terms}
\begin{eqnarray}\label{SSS1}
	\Sigma_{1}&\equiv&-[\partial_{n}(\bm{u}\bm{\cdot}\bm{\nabla}\Omega)]_{w}=-\bm{\xi}\bm{\cdot}\bm{\nabla}_{\pi}\Omega_{w}-\vartheta_{w}\left[\partial_{n}\Omega\right]_{w}\nonumber\\
	&=&-\bm{\xi}\bm{\cdot}\bm{\nabla}_{\pi}\Omega_{w}-\nu^{-1}\vartheta_{w}(\bm{\xi}\bm{\cdot}\bm{\nabla}_{\pi}H_{w})-\vartheta_{w}(\bm{s}_{w}\bm{\cdot}\bm{K}\bm{\cdot}\bm{s}_{w}),
\end{eqnarray}
\begin{eqnarray}\label{SSS2}
	\Sigma_{2}&\equiv&-2\partial_{n}\left[\vartheta\Omega\right]_{w}=-2\vartheta_{w}\left[\partial_{n}\Omega\right]_{w}-2\left[\partial_{n}\vartheta\right]_{w}\Omega_{w}\nonumber\\
	&=&-2\nu^{-1}\vartheta_{w}(\bm{\xi}\bm{\cdot}\bm{\nabla}_{\pi}H_{w})-2\vartheta_{w}(\bm{s}_{w}\bm{\cdot}\bm{K}\bm{\cdot}\bm{s}_{w})-2\left[\partial_{n}\vartheta\right]_{w}\Omega_{w},
\end{eqnarray}
\begin{eqnarray}\label{SSS3}
	\Sigma_{3}&\equiv&\left[\partial_{n}\left(\bm{\omega}\bm{\cdot}\bm{D}\bm{\cdot}\bm{\omega}\right)\right]_{w}=4\left[\partial_{n}R_{D}\right]_{w}\nonumber\\
	&=&-\bm{\xi}\bm{\cdot}\bm{\nabla}_{\pi}\Omega_{w}+2(\bm{\nabla}_{\pi}\bm{\cdot}\bm{\xi})\Omega_{w}-\vartheta_{w}(\bm{s}_{w}\bm{\cdot}\bm{K}\bm{\cdot}\bm{s}_{w}),
\end{eqnarray}
\begin{eqnarray}\label{SSS4}
	\Sigma_{4}&\equiv&\left[\partial_{n}\left(\bm{\omega}\bm{\cdot}\bm{\nabla}T\times\bm{\nabla}s\right)\right]_{w}\nonumber\\
	&=&\left[\partial_{n}\bm{\omega}\right]_{w}\bm{\cdot}\left[\bm{\nabla}T\times\bm{\nabla}s\right]_{w}+\bm{s}_{w}\bm{\cdot}\left[\partial_{n}\left(\bm{\nabla}T\times\bm{\nabla}s\right)\right]_{w},
\end{eqnarray}
\begin{eqnarray}\label{SSS5}
	\Sigma_{5}&\equiv&-\nu\left[\partial_{n}\left(\bm{\nabla}\bm{\omega}\bm{:}\bm{\nabla}\bm{\omega}\right)\right]_{w}\nonumber\\
	&=&-2\nu\left(\bm{K}\bm{\cdot}\bm{\nabla}_{\pi}\bm{s}_{w}\right)\bm{:}\bm{\nabla}_{\pi}\bm{s}_{w}-2\bm{\nabla}_{\pi}\bm{\sigma}\bm{:}\bm{\nabla}_{\pi}\bm{s}_{w}-2\bm{\sigma}\bm{\cdot}\left[\partial_{n}^{2}\bm{\omega}\right]_{w}.
\end{eqnarray}
\end{subequations}
From~\eqref{vorticity_equation}, the second-order wall-normal derivative of vorticity in~\eqref{SSS5} is derived as
\begin{eqnarray}
\left[\partial_{n}^{2}\bm{\omega}\right]_{w}=\nu^{-1}\left(\partial_{t}-\nu\nabla_{\pi}^{2}\right)\bm{s}_{w}
+\nu^{-1}\vartheta_{w}\bm{s}_{w}+\nu^{-1}K\bm{\sigma}-\nu^{-1}\left[\bm{\nabla}T\times\bm{\nabla}s\right]_{w}.
\end{eqnarray}
It is noted that the source term $\Sigma_{3}$ [\eqref{SSS3}] is solely determined by the boundary $R_{D}$ flux.

Particularly, for incompressible flow past a stationary flat wall, $\Sigma_{2}=\Sigma_{4}=0$ and~\eqref{five_source_terms} is simplified as
\begin{subequations}\label{UU}
\begin{eqnarray}\label{UU1}
\Sigma_{1}=-\bm{\xi}\bm{\cdot}\bm{\nabla}_{\pi}\Omega_{w},
\end{eqnarray}
\begin{eqnarray}\label{UU2}
	\Sigma_{3}=-\bm{\xi}\bm{\cdot}\bm{\nabla}_{\pi}\Omega_{w}+2(\bm{\nabla}_{\pi}\bm{\cdot}\bm{\xi})\Omega_{w},
\end{eqnarray}
\begin{eqnarray}\label{UU3}
	\Sigma_{5}=-2\rho^{-1}\bm{\nabla}_{\pi}\bm{\xi}\bm{:}\bm{\nabla}_{\pi}\bm{\nabla}_{\pi}p_{w}-2\mu^{-1}\left(\partial_{t}-\nu\nabla_{\pi}^{2}\right)\bm{\xi}\bm{\cdot}\bm{\nabla}_{\pi}p_{w}.
\end{eqnarray}
\end{subequations}
In~\eqref{UU}, $\Sigma_{1}$ represents the variation of enstrophy along a $\bm{\xi}$-line (or a skin friction line). The sum of $\Sigma_{1}$ and $\Sigma_{3}$ identifies an important characteristic surface quantity, namely $\bm{\xi}\bm{\cdot}\bm{\nabla}_{\pi}\Omega_{w}-(\bm{\nabla}_{\pi}\bm{\cdot}\bm{\xi})\Omega_{w}$, which (multiplied by a constant factor) has been shown as the dominant physical mechanism responsible for the spatiotemporal evolution rate of the wall-normal Lamb dilatation flux at the wall region beneath an energetic quasi-streamwise vortex in a turbulent channel flow~\citep{Chen2023PhysicaD}. $\Sigma_{5}$ is determined by the longitudinal-transverse coupling among $\bm{\xi}$, $p_{w}$ as well as their temporal and tangential derivatives along the boundary.

\section{Conclusions and discussions}\label{Conclusions and discussions}
The present paper provides a necessary theoretical basis for advancing the turbulence theory and computation based on the Lagrangian VGT dynamics to the wall-bounded cases at the fundamental level.
The boundary fluxes of the VGT and its invariants are explicitly unraveled and successfully linked with the established \textit{boundary vorticity and dilatation dynamics} on a stationary rigid wall, where the former can be correspondingly referred to as the \textit{boundary vortex dynamics}.
In this sense, a unified description of the boundary vorticity and vortex dynamics is thereby achieved for the wall-bounded flows.
The main findings and contributions are summarized as follows.

1. On a stationary rigid wall, the whole vorticity dynamics is completely determined by the spin field without the orbital rotation. We find that the VGT $\bm{A}$ at the wall (or the Caswell formula) shows a compatible physical structure as that unveiled by the normal-nilpotent decomposition (NND). Both $\bm{D}_{w}$ and $\bm{\Omega}_{w}$ contain the physical effects due to the spin field, which challenges the existing perspective that $\bm{D}_{w}$ represents the pure deformation of a fluid element.

2. The explicit expressions of the boundary $\bm{A}$-flux are derived, which include both the kinematic and dynamic forms.
They not only provide the required boundary conditions for solving the evolution equation of $\bm{A}$ but also describes its boundary creation mechanisms (thereby improved to be the boundary sources).
The kinematic form emphasizes the crucial roles of the boundary vorticity flux (BVF) and the boundary dilatation flux (BDF) which are expressible in terms of the fundamental surface physical and geometric quantities, naturally leading to the dynamic form. At the high-Reynolds-number limit, the magnitude analysis for an attached boundary layer shows that the surface pressure gradient becomes the dominant boundary source for the viscous boundary $\bm{A}$-flux.

3. The symmetric and anti-symmetric parts of the boundary $\bm{A}$-flux yields the boundary fluxes of $\bm{D}$ and $\bm{\Omega}$. The trace of the boundary $\bm{D}$-flux recovers the expression of the BDF. A concise representation of the boundary $\bm{\Omega}$-flux is obtained with the use of the wedge product, from which the BVF can be easily derived by acting the Hodge star operator. Therefore, a complete theory of the boundary $\bm{A}$-flux is established.

4. The boundary $Q$ flux can be interpreted as the competition among the boundary fluxes of squared dilatation, enstrophy and squared strain-rate. We prove that the boundary $Q$ flux can be created by
the boundary coupling mechanisms of three types: (a) longitudinal-transverse coupling $\left[(\bm{\nabla}_{\pi}\bm{\cdot}\bm{\xi})\vartheta_{w}
-\bm{\xi}\bm{\cdot}\bm{\nabla}_{\pi}\vartheta_{w}\right]$, (b) surface curvature-dilatation interaction $\left[-K\vartheta_{w}^{2}\right]$ and (c) the quadratic interaction $\left[-\bm{\xi}\bm{\cdot}\bm{K}\bm{\cdot}\bm{\xi}\right]$, where $\bm{\xi}\equiv\bm{s}_{w}\times\bm{n}$ is parallel to the skin friction vector. 
Importantly, both the boundary fluxes of enstrophy and squared strain-rate can be created by the coupling between $\bm{\xi}$ and surface pressure gradient $\left[\bm{\xi}\bm{\cdot}\bm{\nabla}_{\pi}p_{w}\right]$, and the other quadratic interaction mechanism $[\bm{s}_{w}\bm{\cdot}\bm{K}\bm{\cdot}\bm{s}_{w}]$, which however, make no net contribution to the boundary $Q$ flux.
In addition, we show that the boundary $R$ flux must be zero as a result of zero boundary value of $Q$.

5. The boundary flux of the third invariant of the strain-rate tensor (namely, the boundary $R_{D}$ flux) is proved to be proportional to the wall-normal derivative of the vortex stretching term, which can be generated by the competition among three mechanisms: (a) the advective variation of the boundary enstrophy along a $\bm{\xi}$-line (or a skin friction line) $\left[-\bm{\xi}\bm{\cdot}\bm{\nabla}_{\pi}\Omega_{w}\right]$, (b) the coupling between the skin friction divergence and the enstrophy $\left[2(\bm{\nabla}_{\pi}\bm{\cdot}\bm{\xi})\Omega_{w}\right]$ and (c) a triple coupling mechanism
among dilatation, vorticity and surface curvature tensor on the boundary $\left[-\vartheta_{w}(\bm{s}_{w}\bm{\cdot}\bm{K}\bm{\cdot}\bm{s}_{w})\right]$. 
The boundary $R_{D}$ flux serves as a source term being responsible for the spatiotemporal evolution rate of the wall-normal enstrophy flux.

Future studies could be devoted to exploring the possible applications of these exact relations in complex flow diagnostics with special focus on revealing the in-depth physical mechanisms of the formation and evolution of complex near-wall coherent structures, detecting the boundary sources of flow noises as well as developing new technologies for significant noise suppression and drag reduction, by virtue of elaborately designed surface configurations. 
Since the surface physical quantities can be calculated from the high-resolution datasets generated through the direct numerical simulations and the real fluid experiments, all the boundary fluxes and the corresponding physical constituents can be quantitatively evaluated, leading to the feasibility of determining the dominated boundary coupling mechanisms in a certain physical problem.
In addition, a generalization to the presence of an arbitrarily moving and deforming boundary could be 
valuable for practical control of near-wall flows.

\section*{CRediT authorship contribution statement} 
\textbf{Tao Chen}: Conceptualization, Methodology, Writing --
original draft, Writing - review \& editing.
\textbf{Jie-Zhi Wu}: Conceptualization, Methodology, Writing --
original draft, Writing - review \& editing. 
\textbf{Tianshu Liu}:  Writing - review \& editing, XXX (experiments).
\textbf{Jie Yao}: XXX (simulations).

\section*{Declaration of competing interest} 
The authors declare that they have no known competing financial
interests or personal relationships that could have appeared to influence
the work reported in this paper.

\section*{Data availability} 
No data is generated for the present study.


\appendix
\setcounter{figure}{0}
\setcounter{table}{0}

\section{Derivation of~\eqref{hh8}}\label{New_App_1}
The boundary $\bm{A}$-flux is evaluated as
\begin{eqnarray}\label{hh1}
	\left[\partial_{n}\bm{A}\right]_{w}=[\partial_{n}(\hat{\bm{\nabla}}_{\pi}+\bm{n}\partial_{n})\bm{u}]_{w}=\bm{\nabla}_{\pi}\left[\partial_{n}\bm{u}\right]_{w}+\bm{n}\left[\partial_{n}^{2}\bm{u}\right]_{w},
\end{eqnarray}
where the no-slip boundary condition ($\bm{u}_{w}=\bm{0}$) has been used. $\hat{\bm{\nabla}}_{\pi}$ is the tangential gradient operator used in the surface vicinity which reduces to ${\bm{\nabla}}_{\pi}$ on the wall. It should be noted that in general, the order of $\hat{\bm{\nabla}}_{\pi}$ and $\partial_{n}$ cannot be arbitrarily exchanged in the presence of a deforming surface. From~\eqref{A_w}, we obtain
\begin{subequations}
	\begin{eqnarray}\label{hh2}
		\left[\partial_{n}\bm{u}\right]_{w}=\vartheta_{w}\bm{n}+\bm{\xi},
	\end{eqnarray}
	\begin{eqnarray}\label{hh3}
		\bm{\nabla}_{\pi}\left[\partial_{n}\bm{u}\right]_{w}
		=(\bm{\nabla}_{\pi}\vartheta_{w})\bm{n}-\vartheta_{w}\bm{K}+\bm{\nabla}_{\pi}\bm{\xi}.
	\end{eqnarray}
\end{subequations}

By applying the identity
\begin{eqnarray}\label{hh4}
	\left[\nabla^{2}\bm{u}\right]_w=\nabla_{\pi}^{2}\bm{u}_{w}-K[\partial_{n}\bm{u}]_{w}+[\partial_{n}^{2}\bm{u}]_{w}=-K[\partial_{n}\bm{u}]_{w}+[\partial_{n}^{2}\bm{u}]_{w},
\end{eqnarray}
it is direct to show that
\begin{subequations}
	\begin{eqnarray}\label{hh5a}
		[\partial_{n}^{2}\bm{u}]_{w}=K[\partial_{n}\bm{u}]_{w}+\left[\nabla^{2}\bm{u}\right]_w=K[\partial_{n}\bm{u}]_{w}+[\bm{\nabla}\vartheta]_{w}-[\bm{\nabla}\times\bm{\omega}]_{w}.
	\end{eqnarray}
	\begin{eqnarray}\label{hh5b}
		\bm{n}[\partial_{n}^{2}\bm{u}]_{w}=K\bm{A}_{w}+\bm{n}[\bm{\nabla}\vartheta]_{w}-\bm{n}[\bm{\nabla}\times\bm{\omega}]_{w}.
	\end{eqnarray}
\end{subequations}

Consider the following orthogonal decomposition for $[\bm{\nabla}\times\bm{\omega}]_{w}$:
\begin{eqnarray}\label{curl_omega}
	[\bm{\nabla}\times\bm{\omega}]_{w}
	=\left(\bm{n}\bm{\cdot}[\bm{\nabla}\times\bm{\omega}]_{w}\right)\bm{n}
	+\bm{n}\times([\bm{\nabla}\times\bm{\omega}]_{w}\times\bm{n}).
\end{eqnarray}
The first term in the right hand side of~\eqref{curl_omega} can be evaluated as
\begin{eqnarray}\label{hh6}
	\bm{n}\bm{\cdot}[\bm{\nabla}\times\bm{\omega}]_{w}=\bm{\nabla}_{\pi}\bm{\cdot}\left(\bm{s}_{w}\times\bm{n}\right)+\bm{s}_{w}\bm{\cdot}\left(\bm{\nabla}_{\pi}\times\bm{n}\right)=\bm{\nabla}_{\pi}\bm{\cdot}\bm{\xi},
\end{eqnarray}
where we have used $\bm{\nabla}_{\pi}\times\bm{n}=\bm{0}$. By virtue of~\eqref{total_BVF}, the second term is rewritten as
\begin{eqnarray}\label{hh7}
	\bm{n}\times([\bm{\nabla}\times\bm{\omega}]_{w}\times\bm{n})=\bm{n}\times\left(\left[\partial_{n}\bm{\omega}\right]_{w}-\bm{K}\bm{\cdot}\bm{s}_{w}\right).
\end{eqnarray}
Therefore,~\eqref{curl_omega} is converted to
\begin{eqnarray}\label{curl_omega2}
	[\bm{\nabla}\times\bm{\omega}]_{w}
	=\left(\bm{\nabla}_{\pi}\bm{\cdot}\bm{\xi}\right)\bm{n}
	+\bm{n}\times\left(\left[\partial_{n}\bm{\omega}\right]_{w}-\bm{K}\bm{\cdot}\bm{s}_{w}\right).
\end{eqnarray}
Combining~\eqref{hh1},~\eqref{hh3},~\eqref{hh5b} and~\eqref{curl_omega2} yields~\eqref{hh8}.

\section{Recovery of boundary vorticity flux}\label{recovery}
For any two vectors $\bm{\xi}$ and $\bm{\eta}$ in the three-dimensional space, it always holds that $\bm{\xi}\wedge\bm{\eta}=\bm{\xi}\bm{\eta}-\bm{\eta}\bm{\xi}$ where $\wedge$ is the wedge product used in the differential geometry~\citep{ChenWH2002}. This implies a concise and self-consistent representation of~\eqref{BOmegaF} as
\begin{eqnarray}\label{BOmegaF2}
	2\left[\partial_{n}\bm{\Omega}\right]_{w}=\bm{n}\wedge\left(K\bm{\xi}+\nu^{-1}\bm{\nabla}_{\pi}H_{w}\right)+\bm{\nabla}_{\pi}\wedge\bm{\xi}.
\end{eqnarray}

We notice that the exterior forms involved in~\eqref{BOmegaF2} can be transformed to their dual vectors by introducing the Hodge star operator $(*)$. For any two vectors $\bm{\xi}$ and $\bm{\eta}$ in the three-dimensional space, it holds that $*(\bm{\xi}\wedge\bm{\eta})=\bm{\xi}\times\bm{\eta}$~\citep{GuoZH1988}. By acting the Hodge star operator on each term of~\eqref{BOmegaF2}, it is direct to show that
\begin{subequations}\label{abc36}
	\begin{eqnarray}\label{a36}
		2*\left[\partial_{n}\bm{\Omega}\right]_{w}=[\partial_{n}*(\bm{\nabla}\wedge\bm{u})]_{w}=[\partial_{n}\bm{\omega}]_{w},
	\end{eqnarray}
	\begin{eqnarray}\label{b36}
		*\left(\bm{n}\wedge\left(K\bm{\xi}+\nu^{-1}\bm{\nabla}_{\pi}H_{w}\right)\right)=\nu^{-1}\bm{n}\times\bm{\nabla}_{\pi}H_{w}+K\bm{s}_{w},
	\end{eqnarray}
	\begin{eqnarray}\label{c36}
*(\bm{\nabla}_{\pi}\wedge\bm{\xi})=\bm{\nabla}_{\pi}\times\bm{\xi}=\bm{s}_{w}\bm{\cdot}\bm{K}-K\bm{s}_{w}-(\bm{\nabla}_{\pi}\bm{\cdot}\bm{s}_{w})\bm{n}.
	\end{eqnarray}
\end{subequations}
Combining~\eqref{BOmegaF2} and~\eqref{abc36} recovers the BVF in~\eqref{BVF_abc} where $K\bm{s}_{w}$ in~\eqref{b36} and $-K\bm{s}_{w}$ in~\eqref{c36} exactly cancel each other.

\section{Boundary fluxes of two traces}\label{BFTrace}
By using~\eqref{A_w} and~\eqref{hh9}, the boundary fluxes of the traces [$tr(\bm{A}^2)$ and $tr(\bm{A}\bm{\cdot}\bm{A}^T)$] are evaluated as
\begin{eqnarray}\label{BQF_T2}
	\frac{1}{2}[\partial_{n}tr(\bm{A}^2)]_{w}&=&[\partial_{n}\bm{A}]_{w}\bm{:}\bm{A}_{w}^{T}\nonumber\\
	&=&\vartheta_{w}([\partial_{n}\vartheta]_{w}+K\vartheta_{w}-\bm{\nabla}_{\pi}\bm{\cdot}\bm{\xi})
	+\bm{\xi}\bm{\cdot}\bm{\nabla}_{\pi}\vartheta_{w}
	+\bm{\xi}\bm{\cdot}\bm{K}\bm{\cdot}\bm{\xi},
\end{eqnarray}
\begin{eqnarray}\label{new2}
	\frac{1}{2}[\partial_{n}tr(\bm{A}\bm{\cdot}\bm{A}^T)]_{w}
	&=&[\partial_{n}\bm{A}]_{w}\bm{:}\bm{A}_{w}\nonumber\\
	&=&\vartheta_{w}\left([\partial_{n}\vartheta]_{w}+K\vartheta_{w}-\bm{\nabla}_{\pi}\bm{\cdot}\bm{\xi}\right)+2K\Omega_{w}\nonumber\\
	& &+\bm{\xi}\bm{\cdot}\bm{\nabla}_{\pi}\vartheta_{w}+\nu^{-1}\bm{\xi}\bm{\cdot}\bm{\nabla}_{\pi}H_{w}.
\end{eqnarray}

\section{{An alternative expression of $R$}}\label{AppendixA}
By applying the identity $tr(\bm{A}^{3})=tr(\bm{D}^{3})+3tr(\bm{\Omega}^{2}\bm{\cdot}\bm{D})$, it follows from~\eqref{EQ3} that
\begin{eqnarray}\label{A1}
	R=\frac{1}{3}\left[\vartheta^{3}-3\vartheta{Q}-tr(\bm{D}^{3})-3tr(\bm{\Omega}^{2}\bm{\cdot}\bm{D})\right],
\end{eqnarray}
which was previously obtained by~\citet{Chong1990}.

It is noticed that the following relations exactly hold:
\begin{subequations}\label{A2ab}
	\begin{eqnarray}\label{A2a}
		-3\vartheta{Q}=-3\vartheta\left(Q_{D}+Q_{\Omega}\right)=-3\vartheta{Q}_{D}-\frac{3}{4}\vartheta\omega^{2},
	\end{eqnarray}
	\begin{eqnarray}\label{A2b}
		-3tr(\bm{\Omega}^{2}\bm{\cdot}\bm{D})=-\frac{3}{4}\bm{\omega}\bm{\cdot}\bm{D}\bm{\cdot}\bm{\omega}+\frac{3}{4}\vartheta\omega^{2}.
	\end{eqnarray}
\end{subequations}

Substituting~\eqref{A2a} and~\eqref{A2b} into~\eqref{A1} yields~\eqref{o3}:
\begin{eqnarray}
R&=&\frac{1}{3}\left[\vartheta^{3}-3\vartheta{Q}_{D}-tr(\bm{D}^{3})-\frac{3}{4}\bm{\omega}\bm{\cdot}\bm{D}\bm{\cdot}\bm{\omega}\right]\nonumber\\
&=&R_{D}-\frac{1}{4}\bm{\omega}\bm{\cdot}\bm{D}\bm{\cdot}\bm{\omega},
\end{eqnarray}
where $-3\vartheta\omega^{2}/4$ in~\eqref{A2a} 
and $+3\vartheta\omega^{2}/4$ in~\eqref{A2b} cancel each other during the derivation.
It is worth mentioning that for incompressible flow ($\vartheta=0$), $\bm{\omega}\bm{\cdot}\bm{D}\bm{\cdot}\bm{\omega}$ and $[-2tr(\bm{D}^{3})-(1/2)\bm{\omega}\bm{\cdot}\bm{D}\bm{\cdot}\bm{\omega}]$ are the source terms being responsible for the rate of change of the enstrophy (${\Omega}\equiv\omega^2/2$) and the squared strain-rate ${S}\equiv tr(\bm{D}^2)$, respectively~\citep{Johnson2024ARFM}. Since $2Q=\Omega-S$ holds for incompressible flow, the difference of these two source terms is equal to $-6R$, being consistent with the restricted Euler model~\citep{Vieillefosse1982,Vieillefosse1984} where $D_{t}Q=-3R$ holds after neglecting the effects of viscosity and pressure anisotropy.

\section{{Wall-normal derivative of vortex stretching term}}\label{AppendixB}
By applying the chain rule of the partial derivative, we obtain
\begin{eqnarray}\label{B1}
\left[\partial_{n}\left(\bm{\omega}\bm{\cdot}\bm{D}\bm{\cdot}\bm{\omega}\right)\right]_{w}=2\left[\partial_{n}\bm{\omega}\right]_{w}\bm{\cdot}\bm{D}_{w}\bm{\cdot}\bm{s}_{w}+\bm{s}_{w}\bm{\cdot}\left[\partial_{n}\bm{D}\right]_{w}\bm{\cdot}\bm{s}_{w}.
\end{eqnarray}

Owing to the kinematic constraint $\bm{\xi}\bm{\cdot}\bm{s}_{w}=0$ and~\eqref{Caswell}, the first term in the right hand side of~\eqref{B1} must be zero. Then, the second term can be decomposed as
\begin{eqnarray}\label{B2}
\bm{s}_{w}\bm{\cdot}\left[\partial_{n}\bm{D}\right]_{w}\bm{\cdot}\bm{s}_{w}
=\bm{s}_{w}\bm{\cdot}\bm{\nabla}_{\pi}\bm{\xi}\bm{\cdot}\bm{s}_{w}-\vartheta_{w}\bm{s}_{w}\bm{\cdot}\bm{K}\bm{\cdot}\bm{s}_{w}.
\end{eqnarray}

By using $\bm{\xi}\times\bm{s}_{w}=2\Omega_{w}\bm{n}$ and the vector identity
\begin{eqnarray}\label{B3}
\bm{\nabla}_{\pi}\times\left(\bm{\xi}\times\bm{s}_{w}\right)
=\bm{s}_{w}\bm{\cdot}\bm{\nabla}_{\pi}\bm{\xi}
-\bm{\xi}\bm{\cdot}\bm{\nabla}_{\pi}\bm{s}_{w}
+\left(\bm{\nabla}_{\pi}\bm{\cdot}\bm{s}_{w}\right)\bm{\xi}
-\left(\bm{\nabla}_{\pi}\bm{\cdot}\bm{\xi}\right)\bm{s}_{w},
\end{eqnarray}
it is deduced that
\begin{eqnarray}\label{B4}
\bm{s}_{w}\bm{\cdot}\bm{\nabla}_{\pi}\bm{\xi}\bm{\cdot}\bm{s}_{w}
=-\bm{\xi}\bm{\cdot}\bm{\nabla}_{\pi}\Omega_{w}+2\left(\bm{\nabla}_{\pi}\bm{\cdot}\bm{\xi}\right)\Omega_{w}.
\end{eqnarray}

Finally, substituting~\eqref{B2} and~\eqref{B4} into~\eqref{B1} yields
\begin{eqnarray}
\left[\partial_{n}\left(\bm{\omega}\bm{\cdot}\bm{D}\bm{\cdot}\bm{\omega}\right)\right]_{w}
=-\left[\bm{\xi}\bm{\cdot}\bm{\nabla}_{\pi}\Omega_{w}-2(\bm{\nabla}_{\pi}\bm{\cdot}\bm{\xi})\Omega_{w}+\vartheta_{w}(\bm{s}_{w}\bm{\cdot}\bm{K}\bm{\cdot}\bm{s}_{w})\right].
\end{eqnarray}

\section*{References}

\bibliography{mybibfile}

\end{document}